\documentclass[superscriptaddress, twocolumn, amsmath, amssymb, aps, prl]{revtex4-2}

\usepackage{graphicx}
\usepackage{dcolumn}
\usepackage{bm}
\usepackage{amsmath}
\usepackage[colorlinks,linkcolor=blue,anchorcolor=blue,citecolor=blue]{hyperref}
\usepackage{color}

\usepackage{epstopdf}
\usepackage{float}

\begin{document}

\title{Near-field Spin Chern Number Quantized by Real-space Topology \\of Optical Structures}

\author{Tong Fu}
\affiliation{Department of Physics, City University of Hong Kong, Tat Chee Avenue, Kowloon, Hong Kong, China}

\author{Ruo-Yang Zhang}
\affiliation{Department of Physics, The Hong Kong University of Science and Technology, Clear Water Bay, Kowloon, Hong Kong, China}

\author{Shiqi Jia}
\affiliation{Department of Physics, City University of Hong Kong, Tat Chee Avenue, Kowloon, Hong Kong, China}

\author{C. T. Chan}
\affiliation{Department of Physics, The Hong Kong University of Science and Technology, Clear Water Bay, Kowloon, Hong Kong, China}

\author{Shubo Wang} 
\email{shubwang@cityu.edu.hk} 
\affiliation{Department of Physics, City University of Hong Kong, Tat Chee Avenue, Kowloon, Hong Kong, China}
\affiliation{City University of Hong Kong Shenzhen Research Institute, Shenzhen, Guangdong 518057, China}

\begin{abstract}
\textcolor{black}{The Chern number has been widely used to describe the topological properties of periodic structures in the momentum space. Here, we introduce a real-space spin Chern number for the optical near fields of finite-sized structures. This new spin Chern number is intrinsically quantized and equal to the structure's Euler characteristic. The relationship is robust against continuous deformation of the structure’s geometry and is irrelevant to the specific material constituents or external excitation. Our work enriches topological physics by extending the concept of Chern number to the real space, opening exciting possibilities for exploring the real-space topological properties of light.}
\end{abstract}

\maketitle


An essential concept in topological physics is the Chern number—an invariant describing the topological properties of dispersion bands in the momentum space. It has been widely applied to study periodic condensed-matter systems with broken time-reversal symmetry, where the Chern number decides the number of chiral edge states at the interface of two distinct systems \cite{1thouless1982quantized,2haldane1988model}. For the periodic systems with fermionic time-reversal symmetry and spin-orbit interaction, the spin Chern number has been introduced to predict the number of helical edge states \cite{4kane2005quantum,5bernevig2006quantum,6sheng2006quantum,8lv2021measurement}. Akin to the condensed matter systems, periodic optical systems can also support topological states described by the Chern number \cite{9haldane2008possible,10wang2009observation,11,12ozawa2019topological} and spin Chern number \cite{12ozawa2019topological,13hafezi2011robust,14bliokh2015quantum,15khanikaev2013photonic,16chen2014experimental,17wu2015scheme}. These photonic topological states can find applications in high-efficiency lasing \cite{18bandres2018topological,19harari2018topological} and robust optical communications \cite{20yang2020terahertz}. 

In addition to the momentum-space topological properties, there is a growing interest in the real-space topological properties of optical systems. Optical fields can exhibit nontrivial topology in the real space, forming knots and links \cite{23dennis2010isolated,24kedia2013tying}, toroids \cite{29shen2021supertoroidal, 30zdagkas2022observation}, and skyrmions \cite{34tsesses2018optical,32du2019deep}. Interestingly, the polarization of optical fields can also generate complex topological configurations such as Möbius strips \cite{25bauer2015observation,35freund2011optical,36peng2022topological,37bauer2016optical}. These real-space topological optical fields can be characterized by some invariants (e.g., skyrmion number) different from the Chern number, and they provide rich degrees of freedom for high-precision light manipulation with potential applications in encoding information \cite{38larocque2020optical}, metrology \cite{39yuan2019detecting}, and sensing \cite{40jia2022chiral}. 

Finding the invariants of topological optical fields is essential to comprehensively understanding the emergence of nontrivial field patterns and singularities. Revealing the relationship between different topological quantities can offer insightful physical pictures for abstract topological concepts. For instance, the Chern number can be interpreted as the winding of the geometric phase on the Brillouin-zone torus, where the geometric phase arises from the evolution of Bloch states \cite{41weng2015quantum}. Geometric phases can emerge in various parameter spaces in addition to the momentum space \cite{42berry1984quantal}. In the real space, the evolution of electromagnetic states can also give rise to geometric phases \cite{43pancharatnam1956generalized,44berry1987adiabatic,45shitrit2013spin,46yin2013photonic}. Is it possible to derive a monopole-type topological invariant similar to the Chern number from the real-space geometric phase? What topological properties are described by this invariant? 

In this Letter, we introduce a new type of spin Chern number based on the geometric phase of optical near fields in finite-sized structures, thus generalizing this important concept from the momentum space to the real space. This spin Chern number characterizes the global topological properties of optical polarization on the structures’ surfaces. Unlike the momentum-space Chern number and other real-space invariants which have no relevance to the real-space topology of optical structures, the spin Chern number here is intrinsically quantized by the genus (i.e., number of “holes”) of optical structures and is guaranteed equal to the Euler characteristic by the Poincaré–Hopf (PH) theorem. This relationship, \textcolor{black}{analytically proved and numerically verified,} exists in general metal structures of arbitrary geometry and is independent of the specific material constituents or external excitations, as long as the structures have smooth surfaces with a small skin depth. 

We first define the new spin Chern number and then apply it to several examples to discuss the physics. A general complex magnetic field in three-dimensional space can be expressed as $\mathbf{H}(\mathbf{r})=\mathbf{e}(\mathbf{r}) H(\mathbf{r})$, where $\mathbf{e}(\mathbf{r})= \mathbf{A}(\mathbf{r})+ i\mathbf{B}(\mathbf{r})$ is the normalized polarization vector with $\mathbf{e}^* \cdot \mathbf{e}=A^2+B^2=1$ and $H(\mathbf{r})=|\mathbf{H}| \mathrm{e}^{i \arg (\mathbf{H} \cdot \mathbf{H}) / 2}$. Here, $\mathbf{A}(\mathbf{r})$ and $\mathbf{B}(\mathbf{r})$ are the major and minor axes of the polarization ellipse, respectively. The spatial variation of $\mathbf{e}(\mathbf{r})$ can generate geometric phases, including the spin-redirection phase \cite{bortolotti1926memories,rytov1938transition,vladimirskii1941rotation} and the Pancharatnam-Berry phase \cite{47bliokh2019geometric}. For the polarization evolution on a closed loop in the real space, the geometric phase can be determined via a path integral over the loop: $\Phi_{\mathrm{G}}=\oint \mathcal{A} \cdot d \mathbf{r}$, where $\mathcal{A}=-i \mathbf{e}^* \cdot(\nabla) \mathbf{e}=-2 \mathbf{B} \cdot(\nabla) \mathbf{A}$ is the Berry connection with Cartesian components $\mathcal{A}_i \equiv-2 \sum_{j=1}^3 B_j \nabla_i A_j$ \cite{47bliokh2019geometric,07d25ae5bd144b1e867f114f44efa238,berry2019geometry}. Equivalently, it can be determined via a surface integral over the area enclosed by the same loop if $\mathcal{A}$ is non-singular in this area: $\Phi_{\mathrm{G}}=\iint \boldsymbol{\Omega} \cdot d \mathbf{S}$, where $\boldsymbol{\Omega}=\nabla \times \mathcal{A}$ is the Berry curvature. Taking the \textcolor{black}{helicity} of the magnetic field into account, one can define a spin Berry connection $\mathcal{A}_{\text {spin}}=\sigma \mathcal{A}$ and a spin Berry curvature $\boldsymbol{\Omega}_{\text {spin}}=\sigma \boldsymbol{\Omega}$ on a given surface $M$, \textcolor{black}{where $\sigma=\text{sign} (\mathbf{s} \cdot \mathbf{n})=\pm 1$ is the local helicity with $\mathbf{s}=\text{Im} [\mathbf{H}^* \times \mathbf{H}]/|\mathbf{H}|^2$ being the normalized local spin density of magnetic field} and $\mathbf{n}$ being the outward unit normal vector of the surface. \textcolor{black}{Note that the helicity here is different from the traditional optical helicity, which is defined as the projection of spin onto the direction of wavevector \cite{tung1985group}.} We define the spin Chern number as:
\begin{equation}
 \begin{aligned}
      \mathbb{C}_{\text {spin}} =\frac{1}{2 \pi} \iint_M \boldsymbol{\Omega}_{\text {spin}} \cdot d \mathbf{S},
 \end{aligned}
\end{equation}\label{1}
where the integral is carried out over the surface $M$. We note that $\mathbb{C}_{\text {spin}}$ differs from the conventional spin Chern number, which is defined by multiplying the helicity globally after the integration of Berry curvature \cite{4kane2005quantum,5bernevig2006quantum,14bliokh2015quantum}. We apply $\mathbb{C}_{\text {spin}}$ to study light scattering by finite-sized metal structures with smooth surfaces. We assume the structures are made of perfect-electric-conductor (PEC), and the effect of material dispersion and loss will be discussed later. All the numerical results are obtained via full wave simulations with a finite-element package COMSOL.

\begin{figure}\label{Fig1}
\centering
\includegraphics[width=\linewidth]{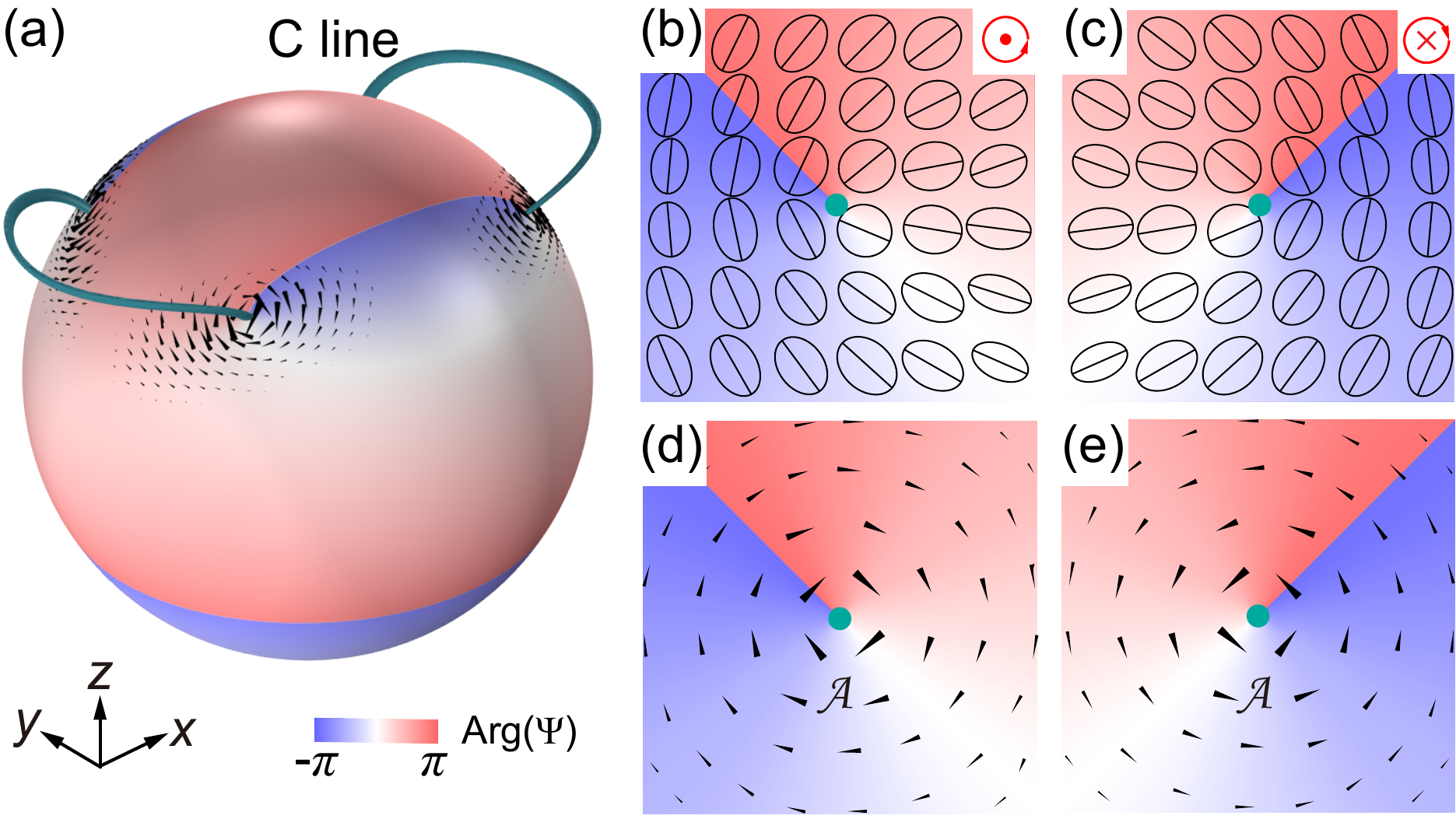} 
\caption{(a) C lines and Berry connection (black arrows) of magnetic field on the PEC sphere excited by a plane wave. The polarization ellipses near the C points with spin pointing (b) outward and (c) inward of the sphere. (d) The Berry connection  corresponding to (b). (e) The Berry connection corresponding to (c). The sphere has a radius $r=400$ nm. The frequency is $f=200$ THz.}
\end{figure}

We consider a PEC sphere under the illumination of a plane wave $\mathbf{H}_{\mathrm{inc}}=\hat{\mathbf{x}} e^{i k z-i \omega t}$. The numerically calculated Berry connection $\mathcal{A}$ on the sphere surface is shown by the black arrows in Fig. 1(a) (see the figure caption for the system parameters). We notice that $\mathcal{A}$ localizes and circulates around four discrete points. These points are C points—polarization singularity at which the field is circularly polarized and the orientation of the polarization major axis $\mathbf{A}$ is ill-defined \cite{48nye1987wave,49berry2004index,50spaegele2023topologically}. The C points correspond to the phase singularities of the scalar field $\Psi=\mathbf{H} \cdot \mathbf{H}=\left(A^2-B^2\right) H^2$, as shown by the color in Fig. 1(a). Since the C points are topological defects of polarization, they can only emerge or annihilate in pairs. Consequently, the surface C points extend into free space to form C lines \cite{36peng2022topological,51garcia2017optical}. Each C line connects a pair of surface C points with opposite helicity or extends to infinity \cite{36peng2022topological,52peng2021polarization}. Figures 1(b,c) and 1(d,e) show the polarization ellipses and Berry connection $\mathcal{A}$, respectively, near the two C points with opposite helicity. The connection $\mathcal{A}$ circulates in opposite directions, indicating its dependence on the \textcolor{black}{helicity} of the magnetic field.

Figure 2(a) shows $\mathcal{A}_{\text {spin}}$  on the sphere (denoted by the black arrows), where the surface color denotes the \textcolor{black}{helicity} $\sigma$. Figure 2(b) shows the value of $\mathbf{n} \cdot \boldsymbol{\Omega}_{\text {spin}}$, which localizes but does not diverge at the C points. This can be understood as follows. If we define another Berry connection for the normalized magnetic field $\mathbf{h} = \mathbf{H}/|\mathbf{H}|$ as $\widetilde{\mathcal{A}}=-i \mathbf{h}^* \cdot(\nabla) \mathbf{h}=\mathcal{A}+\frac{1}{2} \nabla[\operatorname{Arg}(\Psi)]$, which is identical to $\mathcal{A}$ up to a gauge transformation term $\frac{1}{2} \nabla[\operatorname{Arg}(\Psi)]$, the corresponding Berry curvature is $\widetilde{\boldsymbol{\Omega}} = \nabla \times \widetilde{\mathcal{A}} = \nabla \times \mathcal{A} + \frac{1}{2}  \nabla \times \nabla[\operatorname{Arg}(\Psi)] = \nabla \times \mathcal{A} = \boldsymbol{\Omega}_{\text {spin}} / \sigma$. Since $\widetilde{\boldsymbol{\Omega}}$ and $\sigma$ are well-defined and continuous at the C points where $\mathbf{H}$ is a smooth function, $\boldsymbol{\Omega}_{\text {spin}}$ must also be continuous. We apply Eq. (1) to numerically calculate the spin Chern number for the sphere. Remarkably, we find that $\mathbb{C}_{\text {spin}}= 2$. Is the quantized value of $\mathbb{C}_{\text {spin}}$ a coincidence? 

To address the above question, we conduct further simulations for various PEC structures shown in Fig. 2(c-f). The structures are excited by the same plane wave as in Fig. 2(a). For the torus in Fig. 2(c), there are eight C points on the surface connected by four C lines. For the double-torus in Fig. 2(e), twelve C points emerge on the surface, connected by six C lines. In both cases, the \textcolor{black}{helicity} distribution is antisymmetric with respect to the $xoz$-plane and $yoz$-plane. Similar to the case of the sphere, $\mathcal{A}_{\text {spin}}$ and $\boldsymbol{\Omega}_{\text {spin}}$ concentrate near the C points. By numerically integrating $\boldsymbol{\Omega}_{\text {spin}}$ over the surface, we obtain $\mathbb{C}_{\text {spin}}= 0$ for the torus and $\mathbb{C}_{\text {spin}}= -2$ for the double torus. These results imply that the spin Chern number always takes the quantized value identical to the Euler characteristic of the metal structures.

\begin{figure*}\label{Fig2}
\centering
\includegraphics[width=\linewidth]{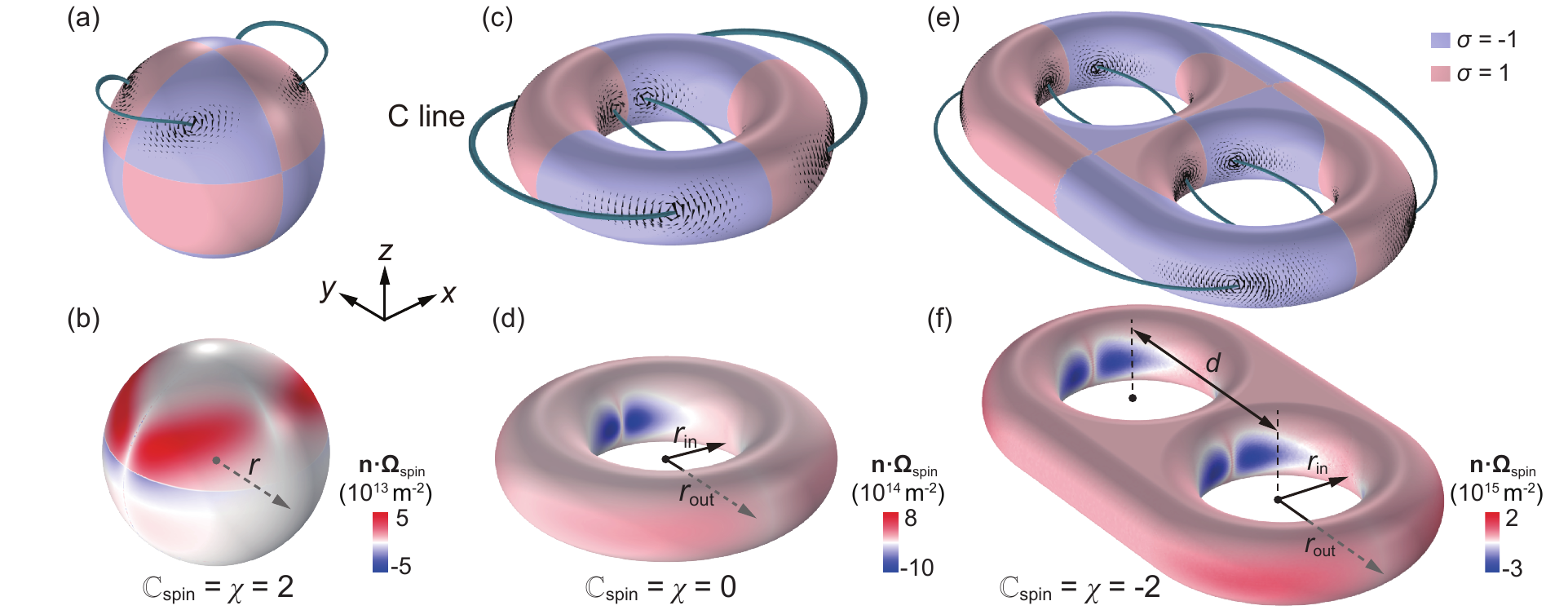} 
\caption{Spin Berry connection (black arrows) on the surface of (a) a sphere, (c) a torus, and (e) a double torus excited by the same plane wave. The surface color denotes the local \textcolor{black}{helicity} $\sigma$. The spin Berry curvature on the surface of (b) the sphere, (d) the torus, and (f) the double torus. The sphere has a radius $r$ = 400 nm. The torus has radii $r_{\text{in}}$ = 110 nm and $r_{\text{out}}$ = 250 nm. The double-torus has $r_{\text{in}}$ = 60 nm, $r_{\text{out}}$ = 120 nm and $d$ = 180 nm.}
\end{figure*}

\textcolor{black}{The mechanism underlying the quantized spin Chern number can be understood with a rigorous analytical proof of its relationship with the topology of optical structures, as we show in the following.}  We divide the metal surface into a set of infinitesimal disks $\left\{D_i\right\}$ each centered at a C point and the exterior region of the disks $M-\sum_i D_i$. Since $\mathcal{A}_{\text{spin}}$ is singular only at the C points, we can apply the Stokes’ theorem to the exterior region to compute the spin Chern number: $\mathbb{C}_{\text {spin}}=\frac{1}{2 \pi} \iint_{M-\sum_i D_i}\left(\nabla \times \mathcal{A}_{\text {spin}}\right) \cdot d \mathbf{S}=-\frac{1}{2 \pi} \sum_i \oint_{\partial D_i} \mathcal{A}_{\text {spin}} \cdot d \mathbf{r}$, where $\partial D_i$ is the boundary of $D_i$ whose positive direction is consistent with $\mathbf{n}$ according to the right-hand rule. Here, we have used $ \iint_{\sum_i D_i} \boldsymbol{\Omega}_{\text {spin}} \cdot d \mathbf{S} = 0$ since $\boldsymbol{\Omega}_{\text {spin}}$ is continuous at the C points. In addition, we have $\mathcal{A}_{\text {spin}}=-2 \sigma A B \mathbf{e}_{\mathrm{B}} \cdot(\nabla) \mathbf{e}_{\mathrm{A}}=-2 \sigma A B\left(\sigma \mathbf{n} \times \mathbf{e}_{\mathrm{A}}\right) \cdot(\nabla) \mathbf{e}_{\mathrm{A}}=-2 A B \mathbf{e}_{\mathrm{B}}^{\prime} \cdot(\nabla) \mathbf{e}_{\mathrm{A}}$, where $\mathbf{e}_{\mathrm{A}} = \mathbf{A} / A$, $\mathbf{e}_{\mathrm{B}} = \mathbf{B} / B$ and $\mathbf{e}_{\mathrm{B}}^{\prime}=\sigma \mathbf{e}_{\mathrm{B}}=\mathbf{n} \times \mathbf{e}_{\mathrm{A}}$ such that $\left\{\mathbf{e}_{\mathrm{A}}, \mathbf{e}_{\mathrm{B}}^{\prime}, \mathbf{n}\right\}$ forms a right-handed basis. \textcolor{black}{Along the lines separating the regions of opposite \textcolor{black}{helicity}, the magnetic field is linearly polarized and the coefficient $2AB$ becomes zero, thereby ensuring the continuity of $\mathcal{A}_{\text {spin}}$. As a result, the Stokes’ theorem can be safely applied.} Near the C points, the coefficient $2AB$ approaches unity, and the spin Berry connection is reduced to $\mathcal{A}_{\text {spin}}=-\mathbf{e}_{\mathrm{B}}^{\prime} \cdot(\nabla) \mathbf{e}_{\mathrm{A}}$. Thus, we have $-\frac{1}{2 \pi} \oint_{\partial D_i} \mathcal{A}_{\text {spin}} \cdot d \mathbf{r}=\frac{1}{2 \pi} \oint_{\partial D_i}\left[\mathbf{e}_{\mathrm{B}}^{\prime} \cdot\left(\nabla \right)\mathbf{e}_{\mathrm{A}}\right] \cdot d \mathbf{r}=\frac{1}{2 \pi} \oint_{\partial D_i} \mathbf{e}_{\mathrm{B}}^{\prime} \cdot d \mathbf{e}_{\mathrm{A}}=I_i$, where $I_i$ is the index of the C point (i. e., winding number of $\mathbf{e}_{\mathrm{A}}$). Finally, we obtain
\begin{equation}\label{2}
\begin{aligned}
    \mathbb{C}_{\text {spin}}(M) & =-\frac{1}{2 \pi} \sum_i \oint_{\partial D_i} \mathcal{A}_{\text {spin}} \cdot d \mathbf{r} \\ 
    & =\frac{1}{2 \pi} \iint_M \boldsymbol{\Omega}_{\text {spin}} \cdot d \mathbf{S} =\sum_i I_i=\chi.
\end{aligned}
\end{equation}
Here, $\chi$ is the Euler characteristic of the structure. The last step corresponds to the application of PH theorem to tangent line fields on smooth manifolds \cite{36peng2022topological,56needham2021visual}, since $\mathbf{A}$ is a line field ($\mathbf{A}$ and $\mathbf{-A}$ denote the same polarization major axis) and the structures' surfaces can be considered smooth manifolds. Equation (2) is the main finding of our work. It shows that the spin Chern number is intrinsically quantized by the topology of the metal structures and is decided solely by the genus $g$ via $\chi=2-2g$. In contrast, the integration of the ordinary Berry curvature always leads to a trivial Chern number $\mathbb{C} = \iint_{M} \boldsymbol{\Omega} \cdot d\mathbf{S} = 0$
regardless of the topology of the structures. \textcolor{black}{This is because $\left\{\mathbf{e}_{\mathrm{A}}, \mathbf{e}_{\mathrm{B}}, \mathbf{n}\right\}$ does not necessarily form a right-handed basis, and $\boldsymbol{\Omega}$ in the regions of opposite handedness cancel each other. The optical spin serves as a hidden degree of freedom dividing the whole surface magnetic field into topologically nontrivial subgroups, akin to the function of fermionic spin that gives rise to the nontrivial momentum-space topology of time-reversal-invariant topological insulators \cite{14bliokh2015quantum,15khanikaev2013photonic,17wu2015scheme}.}

\textcolor{black}{While Eq. (2) is based on rigorous analytical proof, it can be intuitively understood as follows. Consider the case of sphere in Fig. 2 as an example, its spin Chern number can be viewed as the accumulated change of spin Berry phase along latitude circles: $\mathbb{C}_{\text {spin}}=\frac{1}{2 \pi} \int_0^\pi d \theta \partial_\theta \int_{\theta=\text {const}} \mathcal{A}_{\text {spin}} \cdot d \mathbf{r}=\frac{1}{2 \pi}\int_0^\pi d \theta \partial_\theta \Phi_{\text {spin}}^{\mathrm{G}}(\theta)$, where $\theta$ is the polar angle. Since the latitude circles at both $\theta=0$ and $\theta=\pi$ reduce to a point, we have $\Phi_{\mathrm{spin}}^{\mathrm{G}}(0)=\Phi_{\mathrm{spin}}^{\mathrm{G}}(\pi)=0 \bmod 2 \pi$. Therefore, $\mathbb{C}_{\text {spin}}$ must be quantized, and continuous variation of the surface magnetic field will not change $\mathbb{C}_{\text {spin}}$ as long as the field remains nonzero. The field can be continuously varied such that at every point of the sphere it reduces to circular polarization while maintaining the original orientations of the major and minor axes. In this homogeneous circular polarization limit, the normal projection of the spin Berry curvature at each point is exactly identical with the local Gaussian curvature $\Omega_{\text {Gauss}}$ of the sphere (i.e., $\mathbf{n} \cdot \boldsymbol{\Omega}_{\text {spin }} \rightarrow \Omega_{\text {Gauss }}$), and thus $\mathbb{C}_{\text {spin}}$ is determined by the integral of $\Omega_{\text {Gauss}}$, i.e., the Euler characteristic of the structure.}

\begin{figure}[t]\label{Fig3}
\centering
\includegraphics[width=\linewidth]{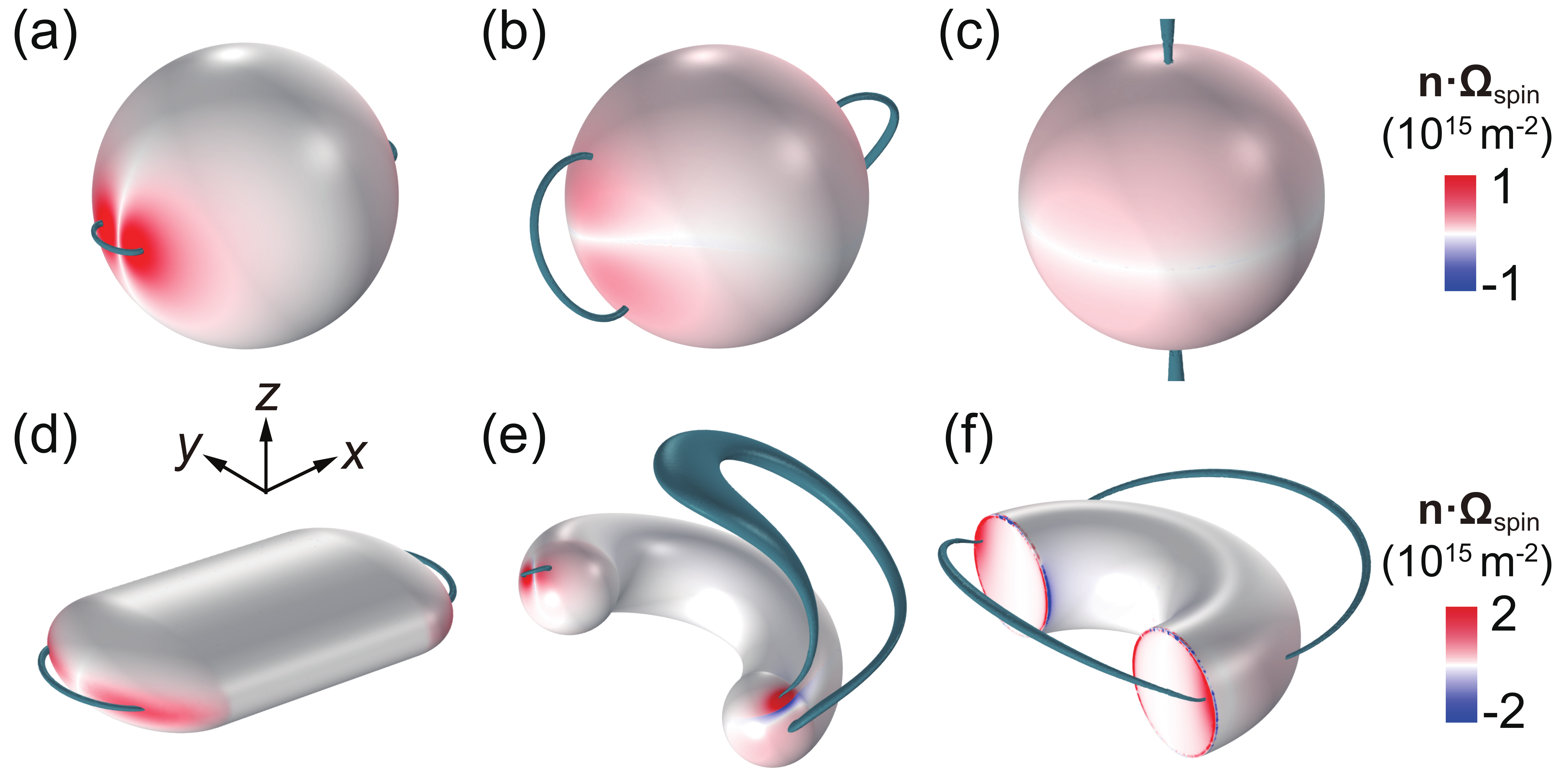}
\caption{The C lines and spin Berry curvature of the PEC sphere excited by different plane waves: (a) $\mathbf{H}_{\mathrm{inc}}=\hat{\mathbf{x}} e^{i k z-i \omega t}$, (b) $\mathbf{H}_{\mathrm{inc}}=(\hat{\mathbf{x}} + i 0.5 \hat{\mathbf{y}})e^{i k z-i \omega t}$, and (c) $\mathbf{H}_{\mathrm{inc}}=(\hat{\mathbf{x}} + i \hat{\mathbf{y}})e^{i k z-i \omega t}$. The C lines and spin Berry curvature in  different geometries without sharp edges [(d) and (e)] and with sharp edges [(f)] excited by the same plane wave as in (a).}
\end{figure}

\begin{table*}
\caption{\label{tab:table1}%
\textcolor{black}{Comparison of topological physics in the momentum space and the real space.}}
\begin{ruledtabular}
{\color{black}\begin{tabular}{ccc}
&Momentum-space topology&Real-space topology (this work) \\
\hline
Physical system&Periodic structures&Finite-sized structures\\
Wave function&Bloch states& Near fields  \\
Topological invariant& (Spin) Chern number & Near-field spin Chern number  \\
Physical property& Protected interface states & Protected interface polarization singularities  \\
Related real-space property  & Symmetry of structures & Topology of structures\\
\end{tabular}}
\end{ruledtabular}
\end{table*}

Equation (2) remains valid for any excitations and continuous deformations of the structure's geometry. Figure 3(a-c) shows the C lines and spin Berry curvature for the PEC sphere excited by plane waves with linear, elliptical, and circular polarizations, respectively, at the same frequency. The different incident waves induce different C lines and $\boldsymbol{\Omega}_{\text {spin}}$. In Fig. 3(a) and 3(b), there are equal number of C points on the surface with different locations, and the associated $\boldsymbol{\Omega}_{\text {spin}}$ is different. In Fig. 3(c), \textcolor{black}{there are only two C points extending from the surface to infinity due to the cylindrical symmetry}, and $\boldsymbol{\Omega}_{\text {spin}}$ is approximately uniform on the surface except at the equator. Numerical calculations confirm that $\mathbb{C}_{\text {spin}}= 2$ in all the three cases. We also verify the effect of geometric deformations, as shown in Fig. 3(d,e), where two different geometries with the same genus $g = 0$ are illuminated by the plane wave $\mathbf{H}_{\mathrm{inc}}=\hat{\mathbf{x}} e^{i k z-i \omega t}$. The C points and $\boldsymbol{\Omega}_{\text {spin}}$ are different in the two cases, but numerical calculations confirm that their spin Chern numbers are both $\mathbb{C}_{\text {spin}}= 2$. The global topology can only be changed by a topological transition of the structure's geometry, e.g., adding/removing holes, or by breaking the conditions of the PH theorem, e.g., adding sharp edges to the structure's surface so that it cannot be considered a smooth manifold. An example is given in Fig. 3(f), where a half torus is excited by the same plane wave as in Fig. 3(d,e). In this case, we obtain $\mathbb{C}_{\text {spin}}= 1.7$ \textcolor{black}{by numerically integrating the spin Berry curvature over the surface, which does not include the contribution from the sharp edges}. The $\mathbb{C}_{\text {spin}}$ is different from the cases in Fig. 3(a-e) due to the sharp edges at which the spin Berry curvature is ill-defined. In fact, the spin Chern number can take arbitrary unquantized values in the presence of sharp edges.

In the above discussions, we have assumed that the structures are made of PEC. The physics also applies to realistic metals with material dispersion and loss, as long as the magnetic field is approximately tangent near the surface. This condition is generally satisfied for various metals at microwave frequencies. At high frequencies, it requires the skin depth of metals to be much smaller than the characteristic geometric dimensions of the structures so that the induced currents localize near the surface and maintain an approximately tangent magnetic field. For dielectric structures, there also exist eigenmodes with tangent magnetic or electric fields near the surface, where similar properties can be found \cite{57bohren2008absorption}. \textcolor{black} {It should be noted that when considering generic perturbations, all stable polarization singularities should be C points \cite{54berry2001polarization,55berry2004polarization}.} \textcolor{black}{The V points with vanished field norm can also emerge on the structure’s surface under certain symmetry, rendering the spin Berry curvature ill-defined at these points. However, the V points are not topologically protected and can split into multiple C points under a generic perturbation, in which case the spin Berry curvature and spin Chern number remain well-defined.} The theory can be naturally extended to the far fields, where the spin Chern number is decided by the topology of momentum sphere \cite{58chen2020line,59horsley2023tutorial,60liu2021topological} and is always $\mathbb{C}_{\text {spin}}= 2$. 

In conclusion, we introduce a new type of spin Chern number for the optical near fields of metal structures with smooth surfaces. The spin Chern number is subtly related to the indices of surface C points and equal to the Euler characteristic of the structures. Thus, it links the topological properties of optical fields and the topological properties of optical structures. The results provide a robust mechanism to manipulate optical near fields via a new degree of freedom, i.e., the topology of structures, which can find applications in high-precision optical metrology, sensing, and imaging. \textcolor{black}{Our work expands the realm of topological physics by extending the concept of monopole-type topological charge from the momentum space to the real space, opening exciting possibilities for exploring the real-space topological properties of light (see Table I for a comparison between the topological physics in the two spaces). The results can be naturally generalized to other types of classical waves such as sound waves and water surface waves, which can bring new insights into these fields.}

The work described in this paper was supported by grants from the Research Grants Council of the Hong Kong Special Administrative Region, China (CityU 11306019 and AoE/P-502/20) and the National Natural Science Foundation of China (12322416 and 11904306).


\bibliography{apssamp}
\bibliographystyle{apsrev4-2}

\pagebreak
\widetext
\begin{center}
\textbf{\large Supplemental Materials for \\
Near-field Spin Chern Number Quantized by Real-space Topology \\ of Optical Structures}
\end{center}

\tableofcontents

\section{NOTE 1. Geometric Phase Associated with C point}

For a closed loop around a C point, the total phase accumulated over the loop can be divided into the geometric phase and the dynamical phase \cite{47bliokh2019geometric}:
\begin{equation}
\Phi = \Phi_{\mathrm{D}}+\Phi_{\mathrm{G}}, \tag{S1}
\end{equation}
where the total phase is related to the magnetic field: $\Phi = -i \oint \frac{\mathbf{H}^* \cdot(\nabla) \mathbf{H}}{|\mathbf{H}|^2} \cdot d \mathbf{r}$. The dynamical phase is determined by the auxiliary scalar field $\Psi=\mathbf{H} \cdot \mathbf{H}$ as   $\Phi_{\mathrm{D}} = - i \oint \frac{\Psi^* \nabla \Psi}{\Psi^2} \cdot d \mathbf{r}$ \cite{47bliokh2019geometric}. The geometric phase is determined by the normalized polarization vector as $\Phi_{\mathrm{G}}= - i \oint\left[\mathbf{e}^* \cdot(\nabla) \mathbf{e}\right] \cdot \mathrm{d} \mathbf{r}$. Since the magnetic field $\mathbf{H}$ is continuous, the total phase must be zero for an infinitesimal loop $\partial D$. In addition, around an arbitrary loop, the dynamical phase is always quantized $\Phi_{\mathrm{D}} = \pi N_D $, where $N_D$ is an integer corresponding to the topological charge of the dynamical phase. Therefore, around the infinitesimal loop $\partial D$, the geometric phase is always quantized:
\begin{equation}
    \Phi_{\partial D}=\Phi_{\mathrm{G}}+\Phi_{\mathrm{D}}=0 \rightarrow \Phi_{\mathrm{G}}=-\Phi_{\mathrm{D}}=-\pi N_D. \tag{S2}
\end{equation}
When the infinitesimal loop encloses a C point with polarization index $I = 1/2$ so that $N_D = \pm 1$, the geometric phase is $\Phi_{\mathrm{G}} = -\Phi_{\mathrm{D}} = \mp \pi$. The geometric phase is related to the local spin (as proven in the main text) and numerically verified in Fig. S1 for the double torus case. There are 12 polarization singularities (C points) for double torus under the excitation of a linearly polarized plane wave, as shown in Fig. S1(a). The arrows in Fig. S1(a) and S1(b) show the Berry connection and spin Berry connection, respectively. The Chern number obtained by the sum of the line integration of the Berry connection is always zero: $\mathbb{C}($double torus$)=-\frac{1}{2 \pi} \sum_{i=1}^{12} \oint_{\partial D_i} \mathcal{A} \cdot d \mathbf{r}=0$, where $\partial D_i$ denote the boundary of an infinitesimal disk $D_i$ centered at the C point. The spin Chern number, on the other hand, is always equal to the Euler characteristic of the geometry: $\mathbb{C}_{\text{spin}}($double torus$)=-\frac{1}{2 \pi} \sum_{i=1}^{12} \oint_{\partial D_i} \mathcal{A}_{\text {spin}} \cdot d \mathbf{r}=-2$. The results of the integrals are summarized in Fig. S1(c).

The vanished Chern number for the double torus is not accidental but is a universal result for structures of any topology. The reason is that, as explained in the main text, $\mathcal{A}$  (Berry connection defined with normalized polarization vector $\mathbf{e}$) and $\widetilde{\mathcal{A}}$ (Berry connection defined with normalized magnetic field $\mathbf{h} = \mathbf{H}/|\mathbf{H}|$) are equivalent up to a gauge, and their corresponding Berry curvatures are identical. Thus, we have 
\begin{equation}
\mathbb{C}(M)=\frac{1}{2 \pi} \iint_M \boldsymbol{\Omega} \cdot d \mathbf{S}=-\frac{1}{2 \pi} \sum_i \oint_{\partial D_i} \mathcal{A} \cdot d \mathbf{r}=-\frac{1}{2 \pi} \sum_i \oint_{\partial D_i} \widetilde{\mathcal{A}} \cdot d \mathbf{r}=\sum_i \frac{\Phi_i}{2 \pi}=0. \tag{S3}
\end{equation}
Here, $\Phi_i$ is the total phase in Eq. (S1) that vanishes for infinitesimal loop $\partial D_i$. As a result, the Chern number of the magnetic field is always zero and has no relation to the topology of the structure.
\begin{figure}[H]
\setcounter{figure}{0}
\renewcommand{\figurename}{FIG.}
\renewcommand{\thefigure}{S\arabic{figure}}
\centering
\includegraphics[width=0.7\linewidth]{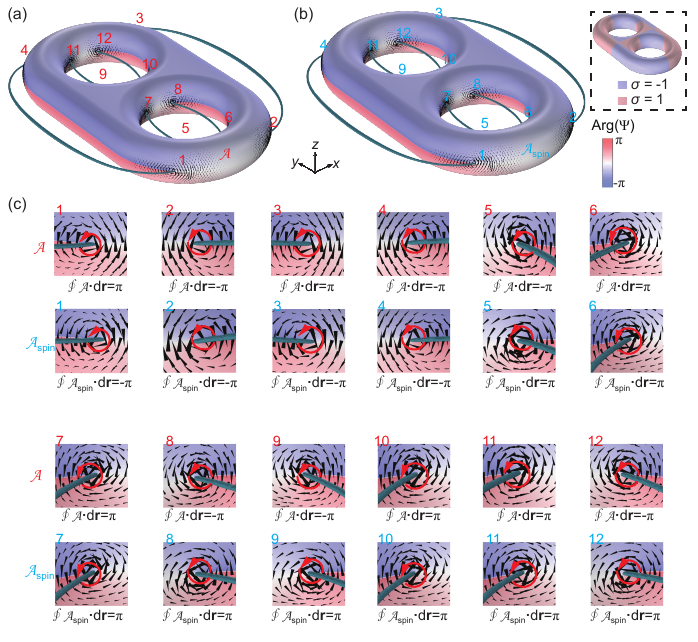} 
\caption{The line integration of Berry connection and spin Berry connection around the C points on a double torus under the incidence of a plane wave propagating in the $z$-direction and with magnetic field linearly polarized in $x$-direction. (a) The distribution of the Berry connection on the surface of the double torus and the C lines. (b) The distribution of the spin Berry connection and the C lines. The inset in the top right corner shows the local helicity defined in the main text. (c) The line integration of the Berry connection and spin Berry connection around an infinitesimal loop enclosing the C points marked in (a) and (b). The background color shows the phase $\text{Arg}(\Psi)$.}
\end{figure}


\section{NOTE 2. Characterizing the Geometric Phase by the Poincarana Sphere}

For paraxial electromagnetic waves with fixed wavevector, the geometric phase, known as the Pancharatnam–Berry (PB) phase \cite{43pancharatnam1956generalized, 44berry1987adiabatic}, can be geometrically described on the Poincaré sphere. For waves with spatially varying wavevector, the spin redirection geometric phase, or Rytov–Vladimirskii–Bortolotti (RVB) phase \cite{bortolotti1926memories,rytov1938transition,vladimirskii1941rotation}, can be geometrically characterized by the unit momentum sphere. In our case, both the polarization and the normal direction of the polarization ellipse vary in space. Therefore, the geometric phase contains both the PB phase and the RVB phase and can be characterized by the Poincarana sphere \cite{47bliokh2019geometric}.
The Poincarana sphere is a unit sphere in the real space. Introducing two-unit vectors $\mathbf{u}_1$ and $\mathbf{u}_2$: $\mathbf{u}_{1,2}= \pm \sqrt{1-\beta^2} \mathbf{e}_{\mathrm{A}}+\beta \mathbf{e}_{\mathrm{s}}$, where $\beta = 2 |\mathbf{A}||\mathbf{B}|$ and $\mathbf{e}_{\mathrm{A}}$ is the unit direction of major axis $\mathbf{A}$ of the polarization ellipse, $\mathbf{e}_{\mathrm{s}}$ is the unit direction of the spin (i.e., the direction of the polarization ellipse). Around a loop in real space, the total geometric phase is equal to half the solid angle swept by the shortest geodesic line connecting the points $\mathbf{u}_1$ and $\mathbf{u}_2$ on the Poincarana sphere. For a closed loop enclosing a C point, the geometric phase described by the Poincarana sphere can be expressed as \cite{47bliokh2019geometric}
\begin{equation}
    \Phi_\text{G} \text{ mod } 2 \pi = (\frac{1}{2} \Sigma +M \pi) \text{ mod } 2 \pi, \tag{S4}
\end{equation}
where $\Sigma$ denotes the total oriented solid angle on the Poincarana sphere. $M$ is a topological number associated with the dynamical phase $M=-N_D \text{ mod } 2$. Here, $N_D$ is the topological number mentioned in NOTE 1. Now we apply this method to characterize the geometric phase in our system with examples.
We consider a sphere excited by a linearly polarized plane wave, as shown in Fig. S2(a). We chose three closed loops marked by 1, 2, and 3 as shown in Fig. S2(b). The loop 1 can be considered to be infinitesimal. Loops 1 and 2 enclose the same polarization singularity C point, while loop 3 encloses two C points of opposite helicity. Hence, we have $M = -N_D =-1$ for loops 1 and 2, and $M = -N_D =0$ for loop 3. The helicity $\sigma$ is shown in Fig. S2(c). The corresponding evolutions of $\mathbf{u}_1$ and $\mathbf{u}_2$ on the Poincarana sphere are shown in Fig. S2(d). For loop 1, the geometric phase given by the Poincarana representation is  $\Phi_\text{G(P)} = (\frac{1}{2} \Sigma +M \pi) \text{ mod } 2 \pi = M \pi = -3.1403$, which agrees with the line integration of Berry connection $\Phi_{\mathrm{G}}=\oint \mathcal{A} \cdot d \mathbf{r} = -3.1421$. For loop 2, the swept solid angle on the Poincarana sphere is $\Sigma = 1.0630$, so the geometric phase is $\Phi_\text{G(P)} = (\frac{1}{2} \Sigma +M \pi) \text{ mod } 2 \pi = 1/2 \times 1.0630 - \pi = -2.6101$, which is also consistent with the direct integration result $\Phi_{\mathrm{G}}=\oint \mathcal{A} \cdot d \mathbf{r} = -2.6161$. For loop 3 enclosing two polarization singularities, we have $M = -N_D = 0$. And, the area swept by $\mathbf{u}_1$ and $\mathbf{u}_2$ forms two closed loops with the opposite direction on the Poincarana sphere. The geometric phase obtained with the Poincarana sphere ($\Phi_\text{G(P)}$) is also identical to the direct integration of $\mathcal{A}$ ($\Phi_\text{G}$), as shown in the right panel of Fig. S2(d).
\begin{figure}[H]
\renewcommand{\figurename}{FIG.}
\renewcommand{\thefigure}{S\arabic{figure}}
\centering
\includegraphics[width=0.8\linewidth]{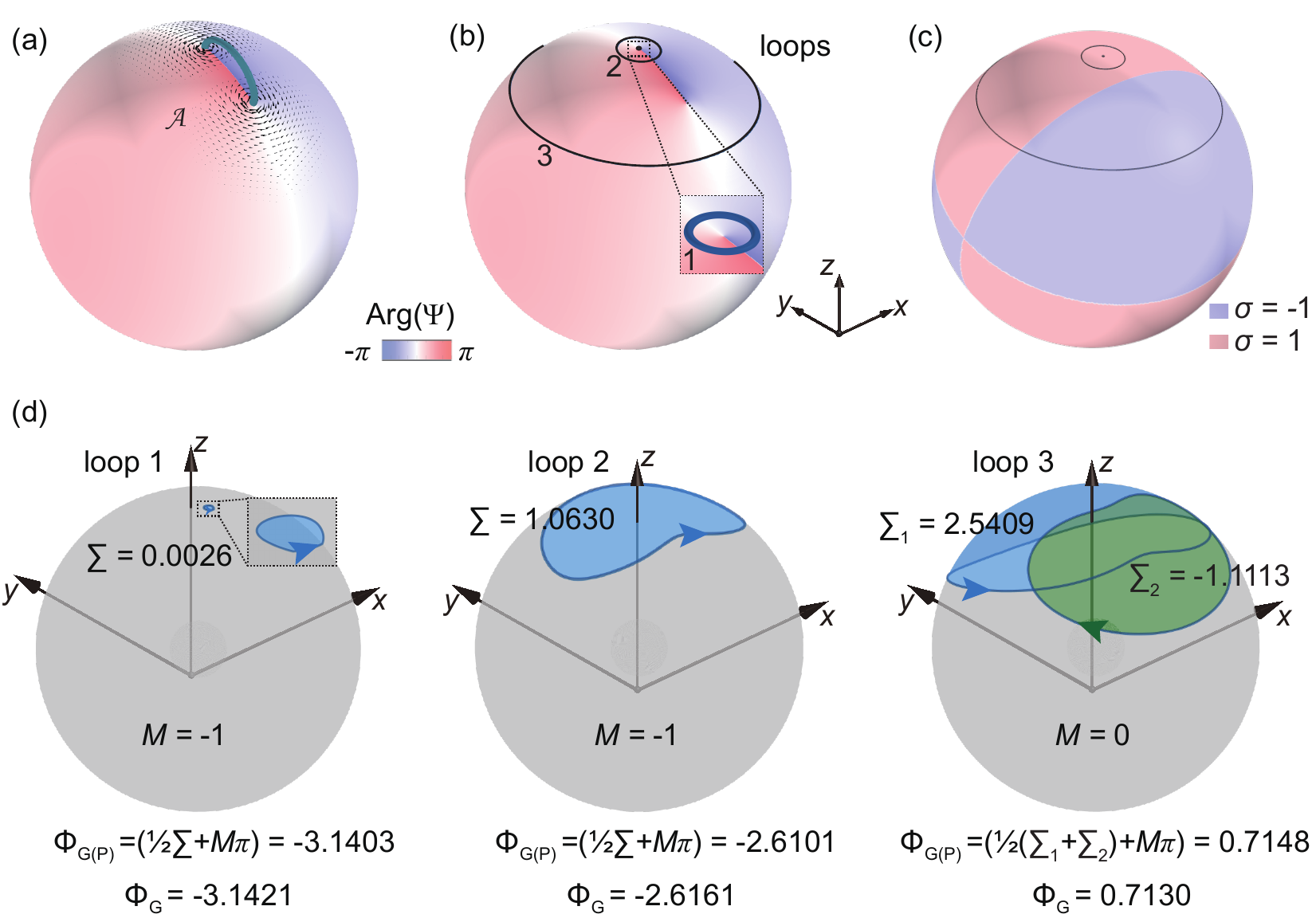} 
\caption{Characterizing the geometric phase by Poincarana sphere. (a) The distribution of the Berry connection and the C line. (b) Three loops are chosen to evaluate the geometric phase. (c) The distribution of local helicity. (d) The swept solid angles on the Poincarana sphere for the three loops in (b). The values below show the geometric phase evaluated with the Poincarana sphere ($\Phi_\text{G(P)}$) according to Eq. (S3) and the geometric phase obtained by direct integration of Berry connection ($\Phi_\text{G}$). The system is excited by a linear polarized plane wave  $\mathbf{H}_{\mathrm{inc}}=(-\hat{\mathbf{y}} + 3 \hat{\mathbf{z}})e^{i k x-i \omega t}$ at $f$ = 200 THz. The sphere is a PEC sphere with a radius of $r$ = 100 nm.}
\end{figure}


\section{NOTE 3. Stokes' Theorem}

Stokes’ theorem can only be applied to the region where the vector field is differentiable and nonsingular everywhere \cite{59horsley2023tutorial}. In the considered scattering system, the Berry connection is well-defined on the structure surface except at the polarization singularity C points. We consider the application of Stokes’ theorem in two cases: 1) the loop does not enclose any C points; 2) the loop encloses C points. As shown in Fig.  S3(a), for the loop $\alpha$ that does not enclose C points, the Berry connections are well-defined everywhere inside the loop (corresponding to the smaller surface area). Thus, Stokes’ theorem can be applied naturally $\Phi_\text{G} = \oint_\alpha \mathcal{A} \cdot d \mathbf{r} = \iint \boldsymbol\Omega \cdot d \mathbf{S}$. For the loop $\beta$ in Fig. S3(a) that encloses a C point, the Berry connection is ill-defined at the C point. To apply Stokes’ theorem, it is necessary to introduce an infinitesimal loop $\beta^{\prime}$ to exclude the singularity, as shown in the inset of Fig. S3(a), and Stokes’ theorem gives $\oint_\beta \mathcal{A} \cdot d \mathbf{r} + \oint_{\beta^{\prime}} \mathcal{A} \cdot d \mathbf{r}=\iint \boldsymbol{\Omega} \cdot d \mathbf{S}$, which is equivalent to carrying out the two path integrals in opposite directions and then taking a sum. Since for 
infinitesimal loop that encloses C points, the geometric phase is quantized (as proved in NOTE 1): $\oint_{\beta^{\prime}} \mathcal{A} \cdot d \mathbf{r} = N_D \pi$. Thus, we have $\oint_\beta \mathcal{A} \cdot d \mathbf{r}+\oint_{\beta^{\prime}} \mathcal{A} \cdot d \mathbf{r}= \oint_\beta \mathcal{A} \cdot d \mathbf{r} + N_D \pi = \iint \boldsymbol{\Omega} \cdot d \mathbf{S}$ . Therefore, the Stokes’ theorem can be expressed as
\begin{equation}
    \oint \mathcal{A} \cdot d \mathbf{r} + N_D \pi = \iint \boldsymbol{\Omega} \cdot d \mathbf{S}, \tag{S5}
\end{equation}
where $N_D$ is the topological charge of the dynamical phase enclosed by the loop. For case 1), no polarization singularity is enclosed, and thus $N_D = 0$. We choose four loops for each case to verify the above equation, as shown in Fig. S3(b). The loops in black all belong to the case 1). The loops in blue all belong to the case 2). Figure S3(c) shows the distribution of the Berry curvature. Figure S3(d) shows the comparisons.

\begin{figure}[H]
\renewcommand{\figurename}{FIG.}
\renewcommand{\thefigure}{S\arabic{figure}}
\centering
\includegraphics[width=0.8\linewidth]{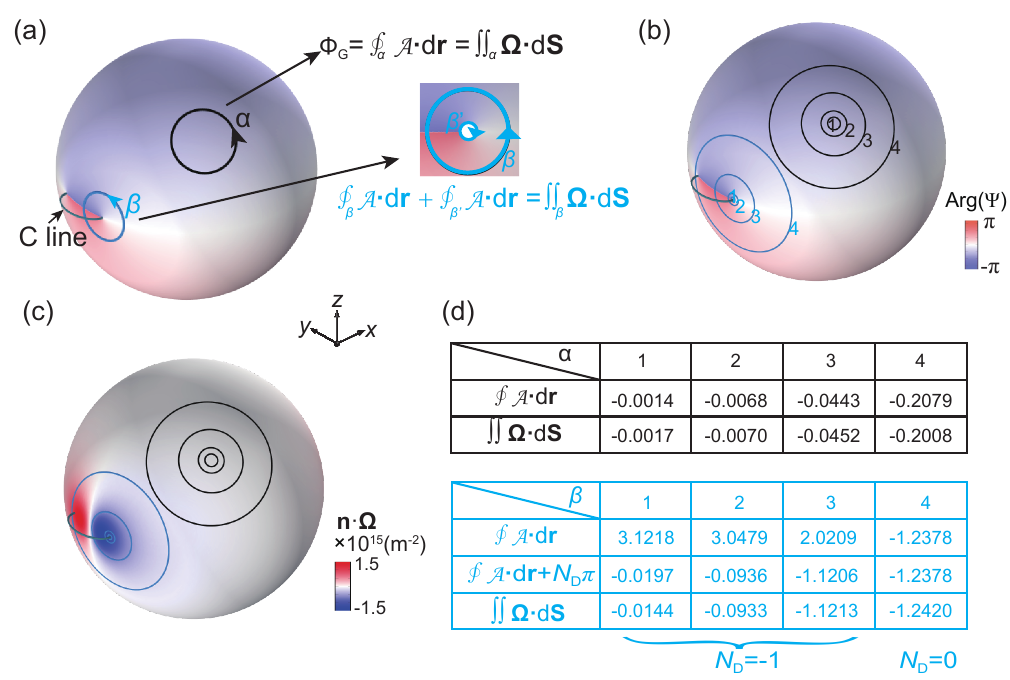} 
\caption{Applying Stokes’ theorem in the considered scattering system. (a) Two types of loops are considered in the application of Stokes’ theorem. The loop $\alpha$ does not enclose C points while the loop $\beta $ encloses a C point. (b) Various loops for verifying the Stokes’ theorem. (c) Distribution of the Berry curvature corresponding to (b). (d) Comparison between the results of surface integral and path integral for varying the Stokes’ theorem in (b). The integration of the Berry connection is along a closed loop in the counterclockwise direction. The integration of Berry curvature is for the smaller area enclosed by the loops. The background color in (a) and (b) shows the dynamical phase $\text{Arg}(\Psi)$.}
\end{figure}


\end{document}